\documentstyle[epsfig]{mn}
\input epsfig.sty
\begin{document}
\hyphenation{bremsstrahlung}
\hyphenation{Ray-mond-smith}
%
%
%
\title[X--ray absorption in the Seyfert 2 galaxy NGC 4507]
 {On the nature of the X--ray absorption in the Seyfert 2  galaxy NGC 4507}
%
%
\author[A. Comastri et al.]{
A. Comastri$^{1}$ \thanks{Andrea Comastri (comastri@astbo3.bo.astro.it)},
C. Vignali$^{2}$,
M. Cappi$^{3,4}$,
G. Matt$^{5}$,
R. Audano$^{2}$, 
H. Awaki$^{6}$, \\ ~ \\
{\LARGE S. Ueno$^{7}$} \\ ~ \\
%
%
\\ $^1$ Osservatorio Astronomico di Bologna, 
via Zamboni 33, I--40126 Bologna, Italy
\\ $^2$ Dipartimento di Astronomia, Universit\`a di Bologna, 
via Zamboni 33, I--40126 Bologna, Italy
\\ $^3$ Istituto per le Tecnologie e Studio Radiazioni Extraterrestre/CNR, 
via Gobetti 101, I--40129 Bologna, Italy
\\ $^4$ The Institute for Physical and Chemical Research RIKEN, 
Wako--shi, Saitama 351--01, Japan
\\ $^5$ Dipartimento di Fisica, Universit\`a degli Studi `Roma Tre', 
via della Vasca Navale 84, I-00146 Roma, Italy
\\ $^6$ Department of Physics, Faculty of Science, Kyoto University, 
Sakyo--ku, Kyoto 606--01, Japan
\\ $^7$ X-ray Astronomy Group, Department of Physics \& Astronomy, University 
of Leicester, Leicester, LE1 7RH, U.K.  
}
%

%
%
\date{Received June 30; accepted October 27, 1997}
%
%
%
\maketitle
\begin{abstract}
We present results of the {\it ASCA} observation of the Seyfert 2 galaxy 
NGC 4507.
The 0.5--10 keV spectrum is rather complex and consists of
several components: (1) a hard X--ray power law heavily absorbed
by a column density of about $3 \times 10^{23}$ cm$^{-2}$,
(2) a narrow Fe K$\alpha$ line at 6.4 keV, (3) soft continuum 
emission well above the extrapolation of the absorbed hard power law,
(4) a narrow emission line at $\sim$ 0.9 keV.
The line energy, consistent with highly ionized Neon 
(Ne\,{\sc ix}), may indicate that the soft  
X-ray emission derives from a combination of resonant scattering and 
fluorescence in a photoionized gas. 
Some contribution to the soft X--ray spectrum from thermal emission,
as a blend of Fe L lines, by a starburst component 
in the host galaxy cannot be ruled out with the present data.
\end{abstract}
%
%
%
\section{Introduction}

Studies of X--ray spectra of Seyfert 2 galaxies above a few keV
(Awaki \& Koyama 1993; Salvati et al. 1997) 
have revealed the presence of highly obscured nuclei 
with power law spectra and Fe K$\alpha$ lines similar to Seyfert 1 objects 
lending further support to the popular `Unification' models for Active 
Galactic Nuclei (AGNs) (Antonucci 1993 and references therein).
In these models the viewing angle explains most of the observed 
differences among Seyfert 1 and Seyfert 2 galaxies in terms of absorption
by circumnuclear matter, possibly a molecular torus. 
If the orientation is such that the line of sight intercepts the 
torus, the Optical/UV radiation including the Broad Lines as well as the
soft X--rays from the nucleus are blocked and the object is classified
as a type 2. 
A fraction of the order of a few percent of the nuclear radiation 
can be detected in scattered light, the scattering medium being a warm 
plasma visible to both the nucleus and the distant observer. 
The origin and the physical state of such a reflecting mirror are 
poorly known at present; this is unfortunate, as the mirror is a key component
of the Unified model (Krolik \& Kallman 1987) and several features originating
from it are expected in X--rays (Krolik \& Kriss 1995; Matt, Brandt \&
Fabian 1996), which should be observable at energies where the torus 
completely blocks the nuclear radiation. However, 
despite extensive studies of Seyfert 2 galaxies with {\it ROSAT} in the soft X--ray 
band (Mulchaey et al. 1993; Turner et al. 1993) 
the relatively low spectral resolution of the PSPC and the weakness
of Seyfert 2 galaxies in the 0.1--3.0 keV energy range hampered detailed 
investigations of the soft component. 

NGC 4507 is a nearby ($z$ = 0.011) barred spiral galaxy, it was  
classified as a SBab(rs)I by Sandage \& Brucato (1979).
Optical spectra of the nucleus show emission lines characteristic of 
a Seyfert 2 type without any detectable broad line component
(Durret \& Bergeron 1986). The relatively high luminosity 
in the O\,$[${\sc iii}$]$ line ($\sim$ 6.5 $\times$ 10$^{41}$ ergs s$^{-1}$; 
Mulchaey et al. 1994), which is though to be a good indicator 
of the luminosity of the Active Nuclei, suggests the presence of a 
powerful source of ionizing radiation. 
NGC 4507 is also a very bright far infrared source with a 60--100 $\mu$m 
luminosity derived from {\it IRAS} fluxes of $\sim$ 10$^{44}$ ergs s$^{-1}$.
In soft X--rays NGC 4507 is rather faint 
and only a marginal detection with the {\it Einstein} IPC 
has been reported (Fabbiano, Kim \& Trinchieri 1992). 
At higher energies (E $>$ 3 keV) it is a bright X--ray source 
and has been observed by several X-ray missions. A
{\it Ginga} observation revealed a hard power law with a flat slope
($\Gamma_{\rm 2-20keV}$ $\sim$ 1.4$\pm{0.2}$), a high column density 
($N_{\rm H}$ $\sim$ 3.7$\pm{0.5}$ $\times$ 10$^{23}$ cm$^{-2}$) and a strong 
iron emission line (EW $\sim$ 400$\pm{100}$ eV) (Awaki et al. 1991; Smith 
\& Done 1996).
OSSE observations (Bassani et al. 1995) showed a steeper 
photon index $\Gamma$  $\sim$ 2.1$\pm{0.3}$ in the 50--200 keV energy range 
in agreement with those of Seyfert 1 galaxies in the same 
energy range (Johnson et al. 1994). 

Here a 40 ksec {\it ASCA} observation of this bright Seyfert 2 galaxy
is presented with the aim of a better understanding of the 
soft--to--medium  X--ray emission. \\
Throughout the paper a Hubble constant $H_{0}$ = 50 Km s$^{-1}$ Mpc$^{-1}$ 
and a deceleration parameter $q_{0}$ = 0 are assumed.

\section{{\it ASCA} observations and data reduction} 

NGC 4507 was observed with the Gas Imaging Spectrometer (GIS) and ths Solid 
State Spectrometer (SIS) onboard the {\it ASCA} satellite 
(Tanaka, Inoue \& Holt 1994) over the period 1994 February 12--13.
The SIS data were obtained using 2--CCD readout mode, where 2 CCD chips are 
exposed on each SIS with the target at the nominal position.
The source position in SIS1 was slightly offset from the nominal value. 
For this reason a fraction of the order of 20--30 per cent of the source flux 
was lost in the gap between the chips.
Following a software--related problem on board {\it ASCA} 
the data collected from the GIS3 were corrupted. As they could not 
be easily recovered, they were therefore excluded from the analysis.
In the following with GIS we refer to GIS2 data only.

All the SIS data were collected in FAINT telemetry mode, which maximises the
CCD spectral resolution. 
The following criteria have been applied for the selection of good 
times intervals: the spacecraft was outside the South Atlantic Anomaly,
the elevation angle from the Earth's limb was $>$ 5 degrees, the minimum 
bright Earth angle was $>$ 25 degrees, the magnetic cutoff rigidity was 
greater than 6 GeV c$^{-1}$ for SIS data and greater than 7 GeV c$^{-1}$ for GIS data.
`Hot' and flickering pixels were removed from the SIS data, and rise-time 
rejection was applied to exclude particle events for the GIS data.
SIS grades 0,2,3 and 4 were considered for the analysis. 
Finally a short period of unstable pointing at the phase of target 
acquisition was removed manually.
After applying the above selection criteria,  25 ks for each 
SIS and 40 ks for the GIS detector were collected.

We fit the GIS and SIS data above 3 keV with an absorbed power law
plus a narrow Gaussian line. The best fit energies 
in the three instruments are in good agreement.
Since the gain of SIS and GIS were within 1 per cent of their nominal
values, the source spectrum can be considered substantially not affected 
by instrumental gain offset.

NGC 4507 is clearly detected in all instruments together with a 
nearby ($\sim$ 6 arcmin) bright (V = 5.8) A0V star.
In the SIS0 the star is detected at the edge of the other CCD chip 
while is barely visible in the SIS1. 
Moreover the gap between the two chips prevents from a significant 
contamination.
In order to estimate the possible contamination in the GIS field 
we have analysed the star spectrum. 
A thermal Raymond--Smith model ($kT \sim$ 1 keV) provides a good fit to the 
data with some evidence of excess flux at higher energies
suggesting that contamination effects between the two sources 
may be relevant. For this reason we have considered in the subsequent GIS
analysis only counts at energies greater than 3 keV.
Circular extraction cells of radius $\sim$ 3.5 arcmin for SIS and $\sim$ 4 arcmin for
GIS centered on NGC 4507 were used, with corresponding background regions
defined in source--free areas of the same CCD chip for SIS and from
calibration background regions (blanksky) for GIS. 

The background subtracted count rates for NGC 4507 are:
0.113$\pm$0.002 cts s$^{-1}$ and 0.081$\pm$0.002 cts s$^{-1}$ in SIS0 and SIS1, respectively
(0.4--10 keV); 0.097$\pm$0.002 cts s$^{-1}$ in GIS in the 3--10 keV band.

\section{Results}

Source plus background light--curves were accumulated for all the instruments
showing no clear evidence for variability. 
GIS and SIS spectra were binned with more than 20 counts per bin
in order to apply $\chi^2$ statistics. The response matrices, effective areas 
and XRT PSF used were those released with the latest version of 
{\sc FTOOLS (3.6)}.
Since the spectral parameters obtained by fitting an absorbed power law
plus a soft component  
to the SIS0 and SIS1 spectra were all consistent at the $\sim$ 90 per cent level
and the residuals from those fits were very similar, we have 
added the two SIS spectra.
In the overlapping energy range (3--10 keV) the GIS spectrum is consistent 
with the SIS results except for a slight mismatch ($<$ 10 per cent) in the relative
normalizations.
In the following the spectral results are referred to the
combined SIS0+SIS1 spectrum simultaneously fitted with the GIS2 one, leaving
the relative normalizations free to vary. Unless explicitly stated 
all the quoted errors correspond to 90 per cent confidence intervals 
for one interesting parameter ($\Delta \chi^2$ = 2.71).

The continuum emission requires at least two components: an heavily absorbed
hard X--ray power law at high energies and a soft component below $\sim$ 3 keV.
However, this model 
does not provide an adequate fit to the data. Two line--like
excesses are clearly visible in the data/model ratio around 6.4 keV, indicative
of iron K$\alpha$ emission (a feature commonly seen in Seyfert galaxies,
Nandra \& Pounds 1994; Mushotzky et al. 1995; Nandra et al. 1997) 
and around 0.9 keV, whose possible origin will be discussed later on. 
Moreover, strong deviations are present in the 1--3 keV region (Fig. 1). 
In the following subsections we provide a detailed description of the spectral
complexity of NGC 4507.

\begin{figure}
\epsfig{figure=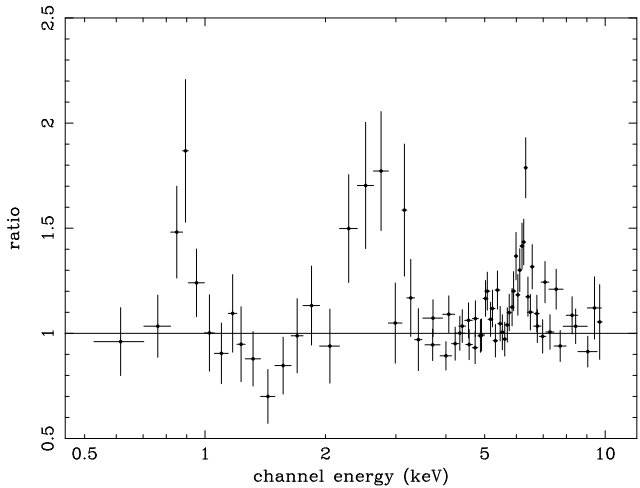,height=7.cm,width=9.5cm,angle=-90}
\caption[]{Data to model ratios in the 0.4--10 keV energy range for a model
consisting of an absorbed power law at high energies plus a power
law component at low energies. The Fe K line at 6.4 keV, a soft line at
0.9 keV and residuals in the 1--3 keV band are clearly visible.}
\label{fig:}
\end{figure}

\subsection{The hard X-ray component}

The hard X--ray spectrum is well fitted by an absorbed ($N_H \sim 10^{23.5}$)
power law model plus a narrow Fe K$\alpha$ line, whose rest energy 
($E = 6.36 \pm 0.03$ keV) indicates emission from neutral matter. 
The line EW (190 $\pm$ 40 eV) is consistent with the 
mean value of Seyfert 1 galaxies (Nandra et al. 1997).
The power law photon index is somewhat dependent on the precise
spectral model chosen for the broad band spectrum. Fitting the data 
in the 3--10 keV energy range we obtain a best fit value of 
$\Gamma = 1.61 \pm 0.20$, which lies at the flatter end of the  
Seyfert 1 photon index distribution, which is characterized by 
an average value of $\Gamma \simeq 1.9$ 
(Nandra \& Pounds 1994; Nandra et al. 1997).
A reflection component has been added to the primary power law
spectrum; however, given the relatively low effective area of {\it ASCA} 
above 6--7 keV, its amplitude is unconstrained by the present data 
and will not be further considered. 
A summary of spectral parameters is reported in Table 1.

\begin{table*}
\caption{3-10 keV Spectral Fits.
$^a$ units of $10^{22}$ cm$^{-2}$; $^b$ in keV; $^c$ in eV; $^{d}$ Total 
$\chi^{2}$ and degrees of freedom.}
\begin{minipage}{100mm}
%
%
\begin{tabular}{lcccc}
\hline\noalign{\smallskip}
 $\Gamma$ & N$_{\rm H}^{a}$ & $E_{line}^{b}$ &  EW$^{c}$ & $\chi^{2}/dof^{d}$  \\
\noalign{\smallskip}
\hline\noalign{\smallskip}
1.78 (1.59--1.98) & 32.4 (30.1--34.7) &  &  & 363.8/278 \\
1.61 (1.41--1.81) & 29.2 (26.9--31.5) & 6.36 (6.34--6.39) & 189 (153--225) &
296.9/276 \\  
\noalign{\smallskip}
\hline
\end{tabular}
%
%
\end{minipage}
\end{table*}

\subsection{The spectrum over the full energy range}

The whole spectrum was then fitted using an absorbed power law plus 
iron line as a baseline model for the high energy spectrum, 
while several different models were fitted to the low energy spectrum
In all cases the absorption by our own Galaxy has been fixed
at the value of $N_{H Gal} = 7.19 \times 10^{20}$ cm$^{-2}$  
(Dickey \& Lockman 1990). The results are reported in Table 2.

\begin{table*}
\caption{0.5--10 keV Spectral Fits. 
$^{a}$ Hard Photon spectral index; 
$^{b}$ Cold absorption column density (units of $10^{22}$ cm$^{-2}$); 
$^{c}$ Soft Photon spectral index; 
$^{d}$ Thermal models temperature (keV); 
$^{e}$ Abundances with respect to the Solar values; 
$^{f}$ Soft Line energy (keV); 
$^{g}$ Soft Line equivalent width (eV); 
$^{h}$ Total $\chi^2$ and degrees of freedom. 
}
\begin{minipage}{175mm}
\begin{tabular}{lcccccccc}
\hline\noalign{\smallskip}
$\Gamma_h^{a}$ & N$_{\rm H}^{b}$ & $\Gamma_s^{c}$ & $kT^{d}$ & $Z/Z_{\odot}^{e}$ & $E_{line}^{f}$ & $EW^{g}$ & 
$\chi^{2}/dof^{h}$ \\
\noalign{\smallskip}
\hline\noalign{\smallskip}
1.49 (1.32--1.67) & 27.8 (26.1--29.9) & 2.43 (2.21--2.65) & \dots & 
 \dots & \dots & \dots & 363.8/321 \\
\smallskip
1.44 (1.27--1.63) & 26.8 (24.6--28.8) & \dots & 1.27 (0.94--1.72) & 
 $<$ 0.03 & \dots & \dots & 380.1/320 \\
\smallskip
1.43 (1.27--1.62) & 26.6 (24.8--28.6) & \dots & 1.25 (1.08--1.65) & 
bremss. &  \dots & \dots & 380.3/321  \\
\smallskip
1.56 (1.40--1.71) & 29.4 (27.7--31.0) & 1.56 ($\equiv$ $\Gamma_h$) 
& 0.71 (0.62--0.79) & 1 (f) & \dots & \dots &  346.3/320 \\
\smallskip
1.52 (1.34--1.71) & 28.4 (26.3--30.5) & 2.14 (1.86--2.40) & \dots 
& \dots & 0.90 (0.88--0.91) & 140 (89--192) & 345.2/319  \\
\smallskip                    
1.74 (1.59--1.88) & 31.2 (29.7--32.7) & 1.74 ($\equiv$ $\Gamma_h$) & \dots & 
\dots & 0.90 (0.88--0.92) & 211 (153--269) & 352.1/320 \\
\noalign{\smallskip}
\hline
\end{tabular}
%
%
\end{minipage}
\end{table*}

Simple thermal models (Bremsstrahlung and Raymond --Smith) for the soft 
component do not provide good fits leaving strong residuals 
at all energies below 3 keV.
A steep $\Gamma = 2.43 \pm 0.22$ power law model for the soft X--ray band
gives instead a better description of the data, with a relative normalization 
with respect to the hard power law of $\sim$ 2 per cent; however, also in this case, 
the fit is rather poor (Table 2), since two remarkable structures 
appear in the residuals: a line--like feature around 0.9 keV and
a waving behaviour in the 1--3 keV range with an `hump' between 2 and 3 keV 
(Fig. 1).
It appears clear that a single power law for the soft X--ray spectrum
which could be interpreted as the fraction of the direct continuum emission
which is scattered into our line of sight by the reflecting mirror
is a too simple an approximation. 
The addition of a narrow line gives a significant improvement 
($\Delta \chi^2 \simeq 19$) with a best fit line energy 
of 0.90 $\pm$ 0.02 keV and EW $\sim$ 140 $\pm$ 50 eV.  

The slope of the low energy power law is steeper than the hard one
and would be inconsistent with a simple scattering model.
However the soft X--ray nuclear spectrum scattered 
into our line of sight could be intrinsically steeper because affected by 
the soft excess frequently observed in Seyfert 1 galaxies.
If the soft photon index is forced to be the same of the hard one
the fit is slightly worse as some flux excess is present in the 0.5--0.7 keV 
region. A possible interpretation of this excess as a blend of
unresolved oxygen lines is discussed below.

An inherent limitation of the {\it ASCA} data is that the complexity of 
the spectrum in terms of absorption and emission lines features 
hampers a precise estimate of the spectral slope.
For this reason in the following we assume that the slope of the 
scattered power law is equal to the hard one ($\Gamma_s \equiv \Gamma_h$).

A combination of a thermal model and a power law for the soft spectrum
clearly improves the fit owing to the greater number of free parameters.
The derived temperature for the thermal component
($kT \sim 0.7 \pm 0.1$ keV) is much lower than the typical values
derived for late type galaxies (kT $\sim$ 3--5 keV; Kim, Fabbiano
\& Trinchieri 1992) such as NGC 4507, but consistent with the characteristic 
temperatures inferred from recent {\it ASCA} observations of 
starburst galaxies (Serlemitsos, Ptak \& Yaqoob 1997).
Leaving the abundances of the Raymond-Smith model free to vary 
there is no improvement in the fit. We note that, 
even with such a complex model, the line--like feature at 0.9 keV
is not completely accounted for (unless Neon abundances are left free to vary) 
and, moreover, the residuals below 
$\sim$ 0.8 keV are steeply increasing towards lower energies (Fig. 2).

\begin{figure}
\epsfig{figure=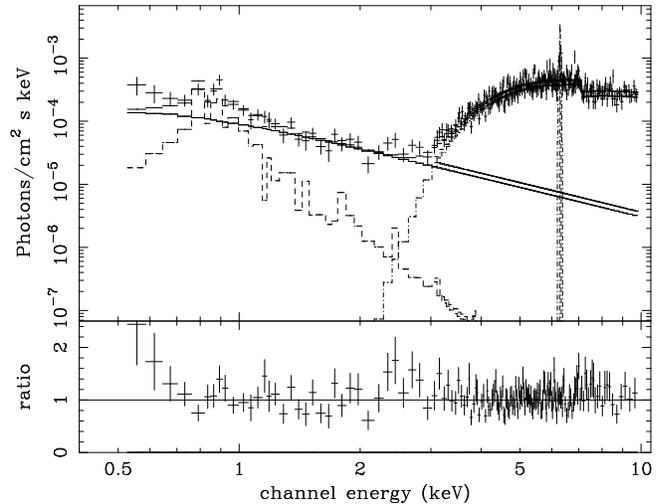,height=7.cm,width=9.5cm,angle=-90}
\caption[]{The NGC 4507 unfolded spectrum, fitted with an absorbed power 
law plus Fe K line at high energies, plus a scattered power law and 
Raymond--Smith thermal emission for the soft energy range (upper panel).
Data to model ratio is shown in the lower panel.}
\label{fig:}
\end{figure}

The shape of the residuals below $\sim$ 3 keV  
and the evidence of a line--like excess at $\sim$ 0.9 keV
suggests, instead (see below),  
that ionized absorption/reflection of the nuclear radiation
by a warm scattering mirror could be important in the 
modelling of the soft X--ray spectrum. 

It should be noted that the putative ionized absorber has significant effects
only on the soft scattered component, while the 
hard X--ray emission is absorbed by almost neutral material
as described in the previous section.

A composite cold plus warm absorber model has been fitted 
to the overall spectrum of NGC 4507 using the 
{\sc ABSORI} model available in {\sc XSPEC 9.0}.
Given the relatively large number of free parameters of the warm absorber
model we have fixed the temperature of the warm material at $T = 10^{5}$ K
(the temperature dependence in this model is very weak in the range
$T = 10^{4.5-5.5}$ K), the iron abundance at the 
solar value and the slope of the primary power law at the same value of
the hard component as is expected if the soft emission is scattered into
our line of sight by the warm mirror.
The fitted parameters are thus the column density $N_W$ of the warm gas
and the ionization parameter $\xi$ which are related by: $\xi = L / n_e R^2$.
A summary of the derived values is reported in Table 3. 
It should be noted that such a model only describes the photoabsorption 
of a background X--ray source, while the case of NGC 4507 is possibly more 
complex owing to the different geometry and physical parameters 
of the scattering region (see below and $\S$ 4.2.1). 
For this reason {\sc ABSORI} should be considered as a first order
approximation for the description of the soft X--ray continuum in NGC 4507.

\begin{table*}
\caption{0.5--10 keV Spectral Fits with warm absorber. 
$^{a}$ Photon spectral index; $^{b}$ Cold absorption column density 
(units of $10^{22}$ cm$^{-2}$); 
$^{c}$ Warm absorption column density (units of $10^{22}$ cm$^{-2}$); 
$^{d}$ Ionization parameter in ergs cm s$^{-1}$; 
$^{e}$ Thermal models temperature (keV); 
$^{f}$ Soft Line energy (keV); $^{g}$ Soft Line equivalent width (eV) with 
respect to the 
unabsorbed continuum; $^{h}$ Total $\chi^2$ and degrees of freedom. }
\begin{minipage}{185mm}
\begin{tabular}{lcccccccc}
\hline\noalign{\smallskip}
$\Gamma_h^{a}$ & N$_{\rm H}^{b}$ & N$_{\rm W}^{c}$ & $\xi^{d}$ & $kT^{e}$ & $E_{line}^{f}$ & $EW^{g}$ &
$\chi^{2}/dof^{h}$ \\
\noalign{\smallskip}
\hline\noalign{\smallskip}
1.89 (1.79--2.02) & 33.6 (32.2--35.1) & 8.5 (4.3--14.2) & 731 (467--1178) &
\dots & \dots & \dots & 369.3/320  \\
\smallskip
1.64 (1.46--1.81) & 31.4 (29.6--33.1) & 8.3 (2.6--20.6) & 520 (220--1194) &
 0.82 (0.69--0.92)  & \dots & \dots & 338.4/318 \\
\smallskip
1.74 (1.62--1.86) & 32.6 (31.2--33.4) & 12.0 (6.1--19.7) & 780 (472--1393) &
\dots & 0.92 (0.90--0.93) & 372 (216--612) & 333.8/318 \\
\noalign{\smallskip}
\hline
\end{tabular}
%
%
\end{minipage}
\end{table*}

With such a model the waving structure in the 1--3 keV range can be almost 
entirely accounted for by the characteristic absorption features of warm gas 
the only remaining feature is a strong line at $\sim$ 0.9 keV.
A highly significant improvement (at $>$ 99.99 per cent according to an F--test) 
has been obtained by adding a narrow gaussian line with EW $\sim$ 370 eV 
and $\sim$ 1.3 keV with respect to the unabsorbed (Fig. 3) and absorbed continuum, 
respectively, the latter value in agreement with the calculations by Netzer (1996). 

\begin{figure}
\epsfig{figure=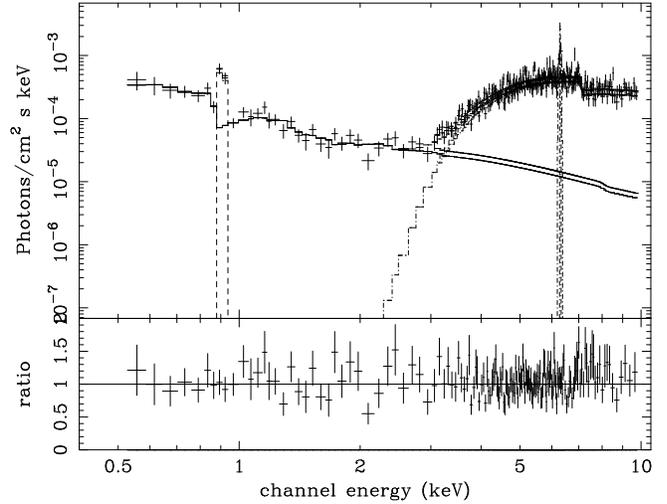,height=7.cm,width=9.5cm,angle=-90}
\caption[]{The best fitting unfolded spectrum. The hard component is fitted 
with an absorbed power law plus Fe K line at 6.4 keV. The soft component
is represented by a warm absorbed power law continuum plus a line at
0.92 keV (upper panel). Data to model ratio is shown in the lower panel.}
\label{fig:}
\end{figure}

The line energy (E = 0.92 $\pm$ 0.02 keV) is consistent with Ne\,{\sc ix} 
and may be 
produced in the photoionized gas with contributions from both the resonant 
scattering and recombination emission (Matt et al. 1996). 
The relative contribution of the two components depends strongly 
on the optical depth of the emitting matter, and will be discuss in the next
section. 
We note that the Ne\,{\sc ix} line is among the most prominent features
predicted in the warm absorber model of Netzer (1996; see his Fig. 4), 
owing to the relatively high abundance of 
Neon and to the large continuum absorption of ionized gas around 0.8--0.9 keV.

An alternative possibility for the line--like emission at 0.9 keV is
in terms of thermal emission from a hot thin plasma,
possibly associated with a starburst component, is also viable.
The improvement with respect to a fit with a warm absorber model 
for the soft spectrum is significant (see Table 3) even if this fit
is not as good as the one with a narrow line at 0.9 keV
($\Delta \chi^2 \sim 5$). 
Leaving the abundances of the Raymond--Smith model free to vary there 
is no improvement in the fit quality.
 
We find a significantly larger ionized column density and ionization parameter with 
respect to the average properties of the warm absorbers in Sey 1s ($<\xi>$ 
$\sim$ 30 ergs cm s$^{-1}$, $<N_{\rm W}>$ $\sim$ a few $\times$ 10$^{21}$ cm$^{-2}$, 
Reynolds 1997). The larger $N_{\rm W}$ is required from the fitting to account 
for the spectral behaviour in the $\sim$ 1.5--3 keV energy range (Fig. 1). 
As a consequence, a greater $\xi$ is needed to account for the data below $\sim$ 1.5 keV. 
We note, however, that these values do not necessarily require a different ionization 
structure between NGC 4507 and Sey 1s', but can be explained with a 
higher inclination angle ({\rm i.e.} a `warm scattering mirror' rather than a `warm absorber'). 
It's interesting to note that even larger values for $N_{\rm W}$ and $\xi$ have been recently 
reported for the Sey 2 galaxy Mkn 3 (Turner et al. 1997).

%
\section{DISCUSSION}

\subsection{The obscured nucleus}

The high energy ($>$ 3 keV) power law slope is relatively flat, but 
consistent with a typical Seyfert 1 spectrum.
The absorption column density, spectral slope and flux level are 
consistent with the previous {\it Ginga} observation, without any evidence
of flux and/or spectral variability over a timescale of about 4 years 
(Awaki et al. 1991; Smith \& Done 1996). On the other hand
the source was a factor 2 brighter in the 2--10 keV band during the 1984
{\it EXOSAT} observation (Polletta et al. 1996).

The observed 2--10 keV flux of the hard power law 
component is 2.1 $\times$ 10$^{-11}$ ergs cm$^{-2}$ s$^{-1}$, 
corresponding to an absorption corrected  luminosity of 
3.7 $\times$ 10$^{43}$ ergs s$^{-1}$,
which is within the range of Seyfert 1 nuclei.

A comparison of the best-fit spectral parameters for the hard component with 
the previous {\it Ginga} values (Awaki et al. 1991, Smith \& Done 1996) suggests 
a possible variation of the Fe $K\alpha$ line intensity and, eventually, of 
the absorption column density. However, given the uncertainties due 
to the lower energy resolution of {\it Ginga} and in the cross-calibration 
of the two instruments, firm conclusions on this issue cannot be drawn. \\
With the present constraints, the Fe $K\alpha$ emission line intensity is 
consistent with transmission trough cold matter with a column density
of a few 10$^{23}$ cm$^{-2}$ (Awaki 1991; Ghisellini, Haardt \& Matt 1994),
though the data do not rule out some contribution from a reflection
component in the continuum and in the line.

\subsection{The soft component}

\subsubsection{Contribution from ionized gas}

The evidence of highly ionized material that leaves significant
imprints on the 0.1-10 keV spectrum of Seyferts galaxies and quasars
is by now widely recognized and, thanks to the {\it ASCA} capabilities,
`warm absorbers' have been clearly detected in several 
Seyfert 1 and quasars (Fiore et al. 1993; Otani et al. 1996; Reynolds 1997).
Several theoretical models have been developed 
to explain the observed features ({\rm e.g.} Netzer 1996 and references therein).

When the ionized gas lies on the line of sight, as for Seyfert 1 objects,
absorption features of the oxygen edges at 0.74 and 0.87 keV 
are usually the most evident characteristics, while the strongest lines 
when observed against the direct continuum have typical equivalent widths of 
a few tens of eV at most, 
so that are difficult to detect with the present detectors.
Much larger equivalent widths for the emission lines 
are expected if the central continuum source is obscured 
as in the case of Seyfert 2 galaxies ({\rm e.g.} Netzer 1996). 

We therefore interpret the observed feature around 0.9 keV we have detected
in the spectrum of NGC 4507 as an emission line (parametrized as a gaussian)
from warm material out of the
line of sight, {\rm i.e.} the same material responsible for the scattering of the
continuum.  In figures 4 and 5
the contour plots of the line energy vs. normalization (with the line
width set to zero) and vs. the width (when permitted to vary) are shown.
The best fit line energy suggests the identification of this feature
with K$\alpha$ emission from 
He-like Neon (0.92 keV). If the reflecting matter would be
optically thin to all processes, the expected EW would be (see Matt et al. 
1996 for the relevant formulae) 
about 6 keV (assuming solar abundances and
a fraction of Ne\,{\sc ix} of 0.4--0.5), resonant scattering accounting
for about 90 per cent of it, to be compared with an observed EW of 
more than one order of magnitude smaller (Table 2 and 3).
However, matter becomes optically thick to resonant
scattering when $N_H\sim$ a few 10$^{19}$ 
and to photoabsorption when $N_H\sim10^{23}$ 
at the edge energy, and at a value five times smaller at the line
energy. Because for these values of the column density Compton scattering is
still optically thin, line EWs are largely dimmed (Matt et al. 1996). 
If the $N_H$ is actually of order of 10$^{23}$, as suggested by
the amount of scattered continuum photons, equivalent widths of the order
of the observed one can be attained.   
A further reduction in the line strength arises from the fact that
the line or, better, the resonant and intercombination lines, is resonantly
trapped in the medium and may be eventually destroyed by photoabsorption
by an Oxygen atom.
The exact value of the line EW depends on several
physical and geometrical details, and its precise evaluation is beyond
the scope of this paper. Note that a strong He-like
oxygen line at 0.57 keV is also to be expected.
Unfortunately, the SIS efficiency at that energy is
rather poor and prevents a detailed study of this feature; in any case,
the obtained upper limit of O/Ne $\sim$ 4--5 is consistent with the
expectation.

\begin{figure}
\epsfig{figure=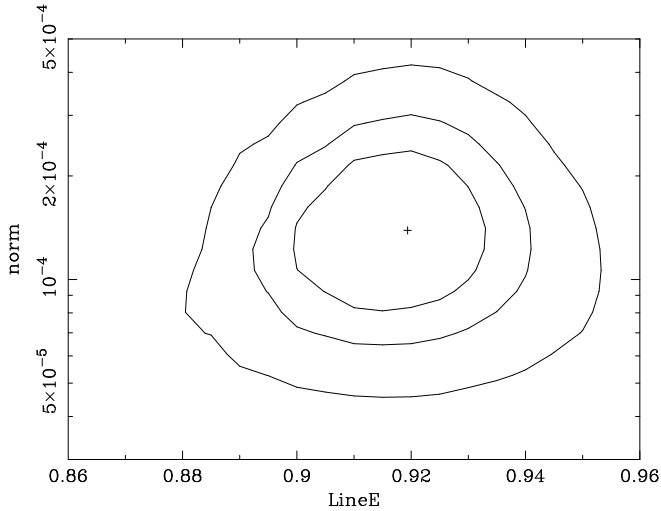,height=7.cm,width=9.5cm,angle=-90}
\caption[]{Confidence contour levels (68, 90 and 99 per cent) for 
two interesting parameters: line energy and flux.}
\label{fig:}
\end{figure}

\begin{figure}
\epsfig{figure=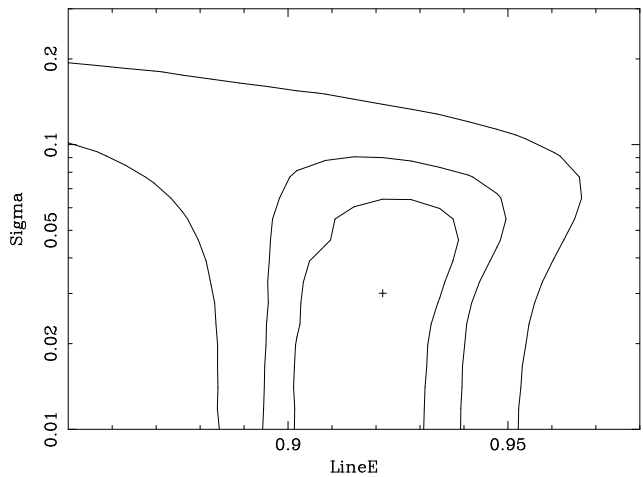,height=7.cm,width=9.5cm,angle=-90}
\caption[]{Confidence contour levels (68, 90 and 99 per cent) for 
two interesting parameters: line energy and width.}
\label{fig:}
\end{figure}

Another possible explanation for the $\sim$0.9 keV feature is in term of
recombination on ground state of completely stripped oxygen, following
a photoionization of an O\,{\sc viii} atom. The 
threshold energy is 0.87 keV, which is inconsistent at the 3$\sigma$ level with
the observed line energy (Fig. 4). 

However, the recombination feature should have
a line--like appearance only if the temperature of the matter is much lower
than the threshold energy; but with a temperature of 10$^6$ K, not impossible
in photoionized plasma, the width of the feature should be not negligible and
still consistent with the observation (Fig. 5), with the centroid energy 
shifted towards higher energy. 
Assuming, as usual, photoionized plasma, the expected EW in the optically
thin case is (for solar abundances and a fraction of O\,{\sc viii} of 0.4-0.5)
about 2 keV. Again, this EW diminishes with the column density, and again
values similar to that observed can be obtained for columns of the order of 
10$^{23}$ cm$^{-2}$. In this case a recombination line at 0.65 keV from 
O\,{\sc viii}, with a similar equivalent width,
is also to be expected. The obtained 90 per cent upper limit to such a line is
about 50 eV.
At these column densities, however, such a line got resonantly
trapped and may be eventually destroyed by photoabsorption by for example
C\,{\sc vi}, which may help explaining the low intensity of this line
with respect to the recombination continuum.
Obviously a final  possibility is 
that both  Ne\,{\sc ix} and the 
O\,{\sc viii} lines can contribute to the observed emission. In fact, 
despite the fact that the ionization potentials of Ne\,{\sc viii} and
O\,{\sc vii} are quite different (as in the last ion the
ionization should involve a K instead than L electron), 
it is possible to have both ions rather abundant at the same time
(see {\rm e.g.} Nicastro et al. 1997).

With the values of the column density of the warm matter as derived by 
both the amount of reflected continuum and the equivalent width of the
0.9 keV feature (whatever its origin), a substantial re--absorption of the
scattered photons is expected. As the matter responsible for both the photons
emission and absorption is the same, the adoption, for the fitting procedure,
of a warm screen in front of the line and continuum emitting region, 
as we have done in the previous section, 
may be not completely 
appropriate to the physical situation under investigation.
A self--consistent grids of models taking into account also emission and 
reflection processes in ionized plasmas, such as those computed using 
e.g. {\sc CLOUDY} or {\sc XSTAR} 
would probably be more appropriate.
We note, however, that the physical picture obtained using  
{\sc ABSORI} plus a Gaussian line at $\sim$ 0.9 keV is in overall good
agreement with much more detailed calculations 
(Krolik \& Kriss 1995; Netzer 1996), and that, in any case, the quality 
of the data is not such 
to permit an unambiguous determination of all the parameters.
 
Interestingly, the derived value of the
absorbing column density is of the same order of that derived for the
reflector, giving a check of our hypothesis that the reflector and the absorber
are one and the same material. The line EW is now somewhat greater, but of the
same order of magnitude,  than in the previous case. 
The best fit ionization structure of the 
absorbing matter suggests that H--like oxygen and neon ions are dominating
over He--like ions, consistent with the O\,{\sc viii} recombination line 
hypothesis for the 0.9 keV feature. 
Observations with 
instruments with higher energy resolution and sensitivity, like AXAF, XMM and
ASTRO--E, are clearly requested to clarify this issue.

\subsubsection {Contribution from the starburst}

Galaxies where the star formation rate greatly exceed the 
average rate of normal galaxies are called starburst galaxies. 
Their optical spectra are characterized by intense narrow emission lines 
due to a recent episode of star formation 
and strong infra--red emission probably due to dust reprocessing
of the radiation from hot stars.
Shock--heated gas is expected to emit in the X--ray band with luminosities
a few order of magnitude lower than those usually observed for Seyfert 
galaxies (see Serlemitsos et al. 1997 for a recent review) 
However when the emission of the AGN is obscured as in the case of 
Seyfert 2 galaxies the starburst component could provide an important 
contribution to the soft X--ray emission.  
An estimate of the expected X--ray luminosity from the starburst component
can be obtained from the far infrared luminosity (David, Jones \& Forman 1992).

Given a FIR luminosity of NGC 4507 of $\sim$ 1.1 $\times$ 10$^{44}$ ergs 
s$^{-1}$ the expected 0.5-4.5 keV luminosity 
is $\sim$ 7.8 $\times$ 10$^{40}$ ergs s$^{-1}$ (equation 1 in David et al. 
1992), while 
the observed luminosity of the soft component in the same energy band is 
$\sim$ 2.2 $\times$ 10$^{41}$ ergs s$^{-1}$, {\rm i.e.} 
about a factor three greater. 
This result suggests that the thermal component 
may account only for a relatively small fraction ($<$ 30 per cent) of the observed
soft X--ray luminosity. 

Previous {\it ROSAT} and {\it ASCA} observations 
of starburst galaxies ({\rm i.e.} Makishima et al. 1994; Moran \& Lehnert 1997)
have shown that the soft X--ray 
emission below 2 keV is frequently extended over a region 
greater than 1 arcmin in extent.
A pointed PSPC observation of NGC 4507 has been retrieved from the 
public archive. The source is rather weak with 
an observed 0.5--2.0 keV flux of $\sim$ 3 $\times$ 10$^{-13}$ ergs cm$^{-2}$ 
s$^{-1}$  consistent with the value derived from our {\it ASCA} analysis
in the same energy range.
The image is consistent with a pointlike source at the PSPC spatial
resolution of $\sim$ 25 arcsec to be compared with the extent of the optical  
image ($\sim 1.3 \times 1.7$ arcmin).   
The lack of any extended emission and the spectral analysis
results indicate that the contribution of a starburst component, 
if present, plays a minor role for explaining the soft X--ray 
emission of NGC 4507.


%
%
\section{Summary}

The main results of the {\it ASCA} observation of the bright Seyfert 2 galaxy
NGC 4507 can be summarized as follows:

$\bullet$ The hard ($>$ 3 keV) power law slope 
($\Gamma \simeq 1.4-1.8$) lies at the flatter--end of the Seyfert 1 photon 
index distribution.
The continuum is strongly absorbed at few keV by a column density
$N_H \simeq 3 \times 10^{23}$ cm$^{-2}$. The iron line
intensity (EW $\simeq$ 190 $\pm$ 40 eV) is consistent with transmission
through cold matter with such a column density.

$\bullet$ The soft X--ray spectrum is rather complex and 
cannot be approximated with a single component. 
A scattered power law plus emission from hot thermal gas   
provide an acceptable description of the soft X--ray continuum, 
but several features are left in the residuals.
A thermal component possibly due to a starburst would in any case 
account for $<$ 30 per cent of the soft X--ray flux.

$\bullet$ Reflection and self--absorption in a photoionized plasma
provide a better description of the overall soft X--ray spectrum.  
A line feature at 0.9 keV is clearly detected, probably due to the 
Ne\,{\sc ix} K$\alpha$ recombination, even if some contribution from
the O\,{\sc viii} recombination continuum cannot be excluded. Note that
the presence of visible soft emission line is Seyfert 2's are expected
(Matt et al. 1996; Netzer 1996), and have
probably already been 
detected by {\it ASCA} in other Seyferts like Mkn 3 (Iwasawa et al. 1994), NGC 4051 
(Guainazzi et al. 1996) and NGC 4388 (Iwasawa et al. 1997), 
as well as in the well known Seyfert 1.5 NGC 4151 
(Leighly et al. 1997).

A broad band observation over the full X--ray domain from 0.1 to several 
tens of keV, like that will be performed by BeppoSAX, 
would allow a better estimate of the relative contribution
of the scattered radiation from the warm gas and the starburst
component. A more detailed study of the warm absorption/reflection
requires good energy resolution and sensitivity; {\it AXAF}, {\it XMM} 
and {\it ASTRO--E} will surely improve significantly our understanding of these phenomena.

\section*{Acknowledgements}

AC acknowledges financial support from the Italian Space Agency under the
contract ASI-95-RS-152. We thank an anonymous referee for useful comments. 
This work has made use of the NASA/IPAC Extragalactic Database (NED)
which is operated by the Jet Propulsion Laboratory, Caltech, under contract
with the National Aereonautics and Space Administration.

%

%

\end{document}